\documentclass[11pt,a4paper]{article}
\usepackage[T1]{fontenc}
\usepackage[utf8x]{inputenc}
\usepackage{graphicx}
\usepackage[margin=1in]{geometry}
\usepackage{rotating}
\usepackage{multirow}
\usepackage{nicefrac}
\usepackage{tabularx}
\usepackage{adjustbox}
\usepackage{tabulary}
\usepackage{booktabs}
\usepackage{setspace}
\usepackage{color}
\usepackage{setspace}
\usepackage{bigstrut}
\usepackage[bottom]{footmisc}
\usepackage[colorlinks=true,urlcolor=blue,linkcolor=blue,citecolor=blue]{hyperref}
\usepackage{threeparttable}
\usepackage{dcolumn}
\newcolumntype{G}{D..{6.4}}
\usepackage{amsmath}
\usepackage{amssymb}
\usepackage{enumerate}   
\usepackage{todonotes}
\usepackage{array}
\usepackage{pdflscape}
\newcolumntype{L}[1]{>{\raggedright\let\newline\\\arraybackslash\hspace{0pt}}m{#1}}
\newcolumntype{C}[1]{>{\centering\let\newline\\\arraybackslash\hspace{0pt}}m{#1}}
\newcolumntype{R}[1]{>{\raggedleft\let\newline\\\arraybackslash\hspace{0pt}}m{#1}}
\newcolumntype{Y}{>{\centering\arraybackslash}X}

\usepackage{siunitx}
\usepackage{newtxtext,newtxmath}

\usepackage{eqparbox}

\usepackage[title,titletoc]{appendix}
\makeatletter

\renewenvironment{appendices}{%
    \begin{oldappendices}%
    \renewcommand{\thefigure}{\ifnum \c@section>\z@ \thesection.\fi\@arabic\c@figure}%
    \@addtoreset{figure}{section}%
    \renewcommand{\thetable}{\ifnum \c@section>\z@ \thesection.\fi\@arabic\c@table}%
    \@addtoreset{table}{section}%
}{%
    \end{oldappendices}%
}\makeatother

\usepackage{natbib}
\let\natbibcitet\citet
\renewcommand\citet{\bibpunct{(}{)}{,}{a}{,}{,}\natbibcitet}
\let\natbibcitep\citep
\renewcommand\citep{\bibpunct{(}{)}{;}{a}{,}{;}\natbibcitep}
\newcommand{\bi}{\begin{itemize}}
\newcommand{\ei}{\end{itemize}}
\newcommand{\be}{\begin{equation}}
\newcommand{\ee}{\end{equation}}
\defcitealias{ieo14}{IEO, 2014}

\long\def\symbolfootnote[#1]#2{\begingroup%
\def\thefootnote{\fnsymbol{footnote}}\footnote[#1]{#2}\endgroup}

\usepackage{sectsty}
\allsectionsfont{\normalsize}

\makeatletter

\@addtoreset{equation}{section}
\makeatother

\usepackage[affil-it]{authblk}

\usepackage{caption}
\usepackage{subcaption}
\makeatletter\let\p@subfigure\thefigure\makeatother

\usepackage{cleveref}
\crefname{chapter}{Chapter}{Chapters}
\crefname{section}{Section}{Sections}
\crefname{subsection}{Section}{Sections}
\crefname{subsubsection}{Section}{Sections}
\crefname{figure}{Figure}{Figures}
\crefname{table}{Table}{Tables}
\crefname{equation}{Equation}{Equations}
\crefname{appendix}{Appendix}{Appendices}

\newcolumntype{d}[1]{D{.}{.}{#1}}

\title{
{\LARGE \textbf{A spatial multinomial logit model for analysing urban expansion}}
}

\usepackage{authblk}

\author[1]{Tam\'{a}s Krisztin\thanks{\textit{Corresponding author}: Tam\'{a}s Krisztin, International Institute for Applied Systems Analysis (IIASA), Schlo{\ss}platz 1, 2361 Laxenburg, Austria. \textit{E-mail}: \href{mailto:philipp.piribauer@wifo.ac.at}{krisztin@iiasa.ac.at}. The research carried out in this paper was supported by funds of the Oesterreichische Nationalbank (Jubilaeumsfond project number: 18043). }}
\author[2]{Philipp Piribauer}
\author[1]{Michael W\"ogerer}

\affil[1]{International Institute for Applied Systems Analysis (IIASA)}
\affil[2]{Austrian Institute of Economic Research (WIFO)}

\date{\vspace{-5ex}}



\begin{document}
\onehalfspacing
\graphicspath{{figs/}}
\maketitle

\begin{abstract}
\noindent
The paper proposes a Bayesian multinomial logit model to analyse spatial patterns of urban expansion. The specification assumes that the log-odds of each class follow a spatial autoregressive process. Using recent advances in Bayesian computing, our model allows for a computationally efficient treatment of the spatial multinomial logit model. This allows us to assess spillovers between regions and across land use classes. In a series of Monte Carlo studies, we benchmark our model against other competing specifications. The paper also showcases the performance of the proposed specification using European regional data. Our results indicate that spatial dependence plays a key role in land sealing process of cropland and grassland. Moreover, we uncover land sealing spillovers across multiple classes of arable land. \\


\end{abstract}

\newpage

\section{Introduction}

Increased urbanisation and expansion of cities as a direct result of economic and population growth, coupled with intensifying climate change, poses a key challenge for policy makers \citep{IPBES2019}. The location choice of new urban developments is of particular importance, as land is a finite resource. Expanding artificial surfaces is both expensive and time consuming to reverse, resulting in long-term impacts on land use and land cover. The conversion of natural habitats to artificial surfaces thus has a direct and potentially irreversible impact on biodiversity \citep{Leclere2018}. On the other hand, if arable land is built up, global food security is threatened and urban expansion might spillover to other types of land use.

Conversion of land to urban surfaces is a decision usually taken by the landowners, which are either regional governments or private land-holders. In an economic framework, this decision is understood as a trade-off between the relative profitabilities of land uses and respective conversion costs \citep{Miller1999}. Potential profits from land ownership are typically assessed using various proxies for land rents \citep{Chakir2017}, while conversion costs rely on the quality of land and national regulations restricting land transformation. Land use change models targeting aggregate administrative levels focus on capturing the outcomes of regional policies \citep{Ay2014}. This is of special importance within the EU, where regional policies (such as the structural funds) are aimed at this level \citep{Alexiadis2013}.

Within a regional econometric framework, the land use expansion decision can be modeled as a random choice, with the  multinomial logit model representing a popular option \citep{Lubowski2008,Chakir2009}. The particular advantage is that a joint modelling of land sealing processes can take into account spillovers across land use classes. When dealing with compositional (shares) data for land use, the multinomial logit random choice model can either be estimated directly from the multinomial logit form \citep{Li2013} or from its log-linearized form \citep{Chakir2017}. While the log-linearized version of the model represents a popular choice due to its ease of transformation, it suffers from the usual problems of log-transformation, namely that frequently land use shares are zero and accommodating these observations inherently biases the estimates.


Spatial dependence, both from unobserved spatially varying variables, as well as contingent on the choice of neighbouring regions, is well documented in the land use choice literature \citep{Chakir2009a,Chakir2013,Li2013}. In a regional econometrics context, a wide number of studies stress the inherent importance of spatial spillovers \citep{LeSage2009}. When estimating models for land use change on a small-scale level, the problem of spatial dependence becomes even more central. Specifically, neglecting to account for spatial autocorrelation may result in severely  biased estimates and erroneous policy conclusions. However, spatial dependence in multinomial logit frameworks has been so far neglected by the spatial econometric literature, with the exception of GMM based approaches \citep{Klier2008}. 


Within this paper our contribution to the existing literature is twofold. First and foremost we present a novel Bayesian approach for capturing spatial dependence among land use changes using a multinomial logit framework. By combining the spatial autoregressive and multionomial logit frameworks, our specification can account for cross-regional and cross-land use class spillovers. The estimation approach builds on recent advances in Bayesian modelling of logit type models \citep{Krisztin2019} and employs latent P\'{o}lya-Gamma distributed variables. We demonstrate the virtues of our approach in a series of Monte Carlo studies. Our second contribution is a novel examination land use change processes on a regional pan-European level. For this we rely on an extensive dataset of land use changes to assess the share of urban gains originating from cropland, grassland, forest and other fallow land. Our framework allows us to shed light on the small-scale spatial dynamics of land sealing processes in European regions. 

The remainder of the paper is structured as follows. Section 2 outlines the theoretical model of urban expansion, as well as its multinomial logit variant. Section 3 focuses on the estimation framework. Section 4 presents the Monte Carlo benchmarks of the proposed econometric estimation approach. Section 5 presents the results for urban expansion in Europe. Section 6 concludes.

\section{A spatial autoregressive multinomial logit model}

In this paper, we estimate an econometric model which aims at  explaining the choice of land buyers (both public and private) for the purpose of converting it to urban, artificial surfaces in $N$ regions. In a given region $i$ (with $i = 1,...,N$), land buyers may acquire land from $J$ different land uses. In our case these are cropland, grassland, forest, and other natural land. Within a region the buyers are assumed to be price taker and their choices are assumed to be homogeneous and risk-neutral. 

In an economic sense this constitutes a profit maximization problem of land buyers \citep{Lubowski2008,Miller1999}, which is directly dependent on the associated profits and costs of the converted land. In addition, to account for the expected net present value of rents from urban land use and the respective conversion costs, land buyers also face the opportunity costs of alternatives usages. 

Such frameworks have been adopted among others by \cite{Lubowski2008}, \cite{Chakir2009a}, and \cite{Li2013}. For estimation of parameters relating to observed buyers' choices, the profit maximization problem can be formulated within a multinomial limited dependent variable framework. Let $y_{ij}$ be the observed share of urban expansion from land use $j$ relative to the total urban expansion in region $i$.\footnote{When land use specific observations $y_{ij}$ are shares, a popular choice for estimating the multinomial model is to apply a log-linear transformation, where the dependent variables correspond to $\log(y_{ij}/y_{iJ})$ and perform standard regression analysis \citep[see, for example][]{Chakir2017,Chakir2009}. The main drawbacks of this approach are twofold. First, Jensen's inequality states that the expectation of a logarithm is not equal to the logarithm of the expectation. Therefore, log-linearization inherently introduces a bias in the estimated slope coefficients. Second, in empirical applications frequently a large number of observed choices $y_{ij}$ are equal to zero, thus necessitating either a censoring of observations or adding a constant to all observations, both of which have been demonstrated to lead to substantial bias.} Econometric estimation thus concerns itself with modelling the probability of observing $y_{ij}$. Within the multinomial logit framework, this probability can be modelled as a function of choice specific log-odds $\mu_{ij}$, weighted by the sum of log-odds over all choice alternatives $\mu_{ij'}$ ($j'=1,\dots,J$):
\begin{align}
p(y_{ij}) = \frac{\exp \mu_{ij}}{\sum_{j'=1}^J \exp \mu_{ij'}}. \label{eq:mnl_model}
\end{align}
In the standard non-spatial multinomial framework $\mu_{ij}$ is specified as a function of $k$ explanatory variables, with corresponding choice-specific slope coefficients, which are to be estimated. The explanatory variables correspond to the expected rents and conversion costs with respect to land use $j$.


Spatial dependence among  log-odds $\mu_{ij}$ in Eq. (\ref{eq:mnl_model}) involves the assumption that the choices of urban land buyers do not solely depend on rent and conversion costs in their own region $i$, but also on other regions' characteristics as well. This assumption implies that the probability of observing a land use choice in region $i$ also depends on land use choices of all other regions. This assumption is based on the spatial nature of land expansion: before  construction, investors typically scope multiple investment opportunities, which might not be contiguous, but located across  regions in spatial proximity to each other. 

Following the spatial econometric literature, such dependencies can be incorporated by imposing an exogenous neighbourhood structure through a non-negative and row-stochastic spatial weight matrix. Let $\boldsymbol{W}$ be such an $N \times N$ spatial weight matrix. Two regions $i$ and $i'$ are assumed to be neighbours of $w_{ii'} > 0$, otherwise $w_{ii'}=0$. No region is a neighbour to itself, thus $w_{ii} = 0$.

The resulting spatial autoregressive (SAR) multinomial logit model can be expressed as:
\begin{align}
\boldsymbol{\mu}_j&=\rho_j\boldsymbol{W}\boldsymbol{\mu}_j+\boldsymbol{X}\boldsymbol{\beta}_j + \boldsymbol{\varepsilon}_j \notag\\
\boldsymbol{\mu}_j &= \boldsymbol{A}^{-1} ( \boldsymbol{X} \boldsymbol{\beta}_j  + \boldsymbol{\varepsilon}_j)  \label{eq:MNL_SAR_logit}, 
\end{align}
with $\boldsymbol{A}_j^{-1} = (\boldsymbol{I}_N - \rho_j\boldsymbol{W})^{-1}$ where $\boldsymbol{I}_N$ denotes an $N \times N$ identity matrix. The $N\times K$ matrix $\boldsymbol{X}=[\boldsymbol{x}_1,\dots,\boldsymbol{x}_K]$ collects the $K$ vectors of explanatory variables and $\boldsymbol{\beta}_j$ denote the respective $K\times 1$ vector of slope parameters related to choice $j$. The $N\times 1$ vector $\boldsymbol{\varepsilon}_j$ contains independently and identically Gaussian distributed disturbance terms, with zero mean and $\sigma_j^2$ variance.  The (scalar) parameter $\rho_j$ measures the strength of spatial autocorrelation for land use class $j$, with sufficient stability condition $\rho_j\in(-1,1)$, where positive (negative) values of $\rho$ indicate positive (negative) spatial autocorrelation. Note, that the model allows for different $\rho_j$ across land use classes.\footnote{With the identifying restriction that the spatial autocorrelation coefficient associated with the $J$-th land use class $\rho_J=0$.} In the absence of spatial autocorrelation ($\rho_1 = ... = \rho_J =0$), the model framework collapses to a classical multinomial logit setup. 

In such a spatial autoregressive (SAR) model specification, the $N\times 1$ vector of choice-specific log-odds $\boldsymbol{\mu}_j=[\mu_{1j},\dots,\mu_{Nj}]'$ thus also depend on the characteristics of other regions in the sample. Spatial dependence is introduced by the spatial multiplier $\boldsymbol{A}_j^{-1}=(\boldsymbol{I}_N-\rho_j\boldsymbol{W})^{-1}=\sum_{r=0}^{\infty} \rho_j^r \boldsymbol{W}^r$.\footnote{It is worth noting that the standard SAR model can be extended to more flexible spatial econometric model specifications in a straightforward way. Specifically, one may additionally include spatially lagged explanatory variables, resulting in a so-called spatial Durbin model (SDM) specification (see, for example \citealt{LeSage2009}). A similar extension is presented in the empirical exercise.} 

A core implication of the SAR modelling framework is that a change in the explanatory variables associated with region $i$, would not only result in changes of the observed shares $y_{ij}$ in the own region, but in other regions as well. Through the nature of the multinomial logit model, where marginal impacts to one choice $j$ also affect the shares of all other choices, this implies that in a spatial dependent setting marginal impacts of $y_{ij}$ have spillover effects over regions and choices as well.

Note, that through the normally distributed residual error vector $\boldsymbol{\varepsilon}_j$, the residuals of the model in Eq. (\ref{eq:MNL_SAR_logit}) are effectively decomposed into two components: first, the heteroscedastic errors arising from the logistic model in Eq. (\ref{eq:mnl_model}) and second, the normally distributed error term $\boldsymbol{\varepsilon}_j$ with $\sigma_j^2$ variance. Spatial dependence in the errors is captured through the latter. Similar to spatial autoregressive variants of standard probit \citep[][]{LeSage2011} and logit models (\citealt{Krisztin2019}), innovation variances $\sigma_j^2$ are restricted to unity, in order to identify the logistic errors.

\section{Estimation strategy}

We propose a Bayesian estimation strategy for the SAR multinomial logit model, which builds on the idea of introducing a latent variable in order to facilitate the estimation of the multinomial logit likelihood. This estimation strategy has been widely employed in recent Bayesian econometric literature for tackling models featuring non Gaussian distributions \citep[see e.g.][]{Fruhwirth-Schnatter2012,Fruhwirth-Schnatter2009}. To illustrate the core problem, consider the likelihood of the multinomial logit model in Eq. (\ref{eq:mnl_model}):
\begin{align}
 \prod_{i=1}^N \prod_{j=1}^J \frac{\left( \exp \mu_{ij} \right)^{y_{ij}}}{\sum_{j'=1}^J \exp \mu_{ij'}}.
\end{align}
Note, that the likelihood contribution of observation $i$ relies not only on $\mu_{ij}$, but on the log-odds of making other choices as well. This well-known non-linearity in the likelihood greatly complicates the estimation of the unknown slope and spatial autoregressive coefficients. 

Within a Bayesian framework the focus of estimation frequently lies mainly on finding conditional posterior distributions for the parameters of interest. In fact, assuming suitable priors $p(\boldsymbol{\beta}_j)$, the conditional posterior of $\boldsymbol{\beta}_j$ can be expressed conditional on all other slope coefficients $\boldsymbol{\beta}_{-j}$ and $\rho$ \citep[see][]{Holmes2006}:
\begin{align}
p\left(\boldsymbol{\beta}_j|\boldsymbol{\beta}_{-j}, \rho\right) &= p\left(\boldsymbol{\beta}_j\right) \prod_{i=1}^N \left( \frac{\exp \eta_{ij}}{1 + \exp \eta_{ij}} \right)^{y_{ij}} \left( \frac{1}{1 + \exp \eta_{ij}} \right)^{1-y_{ij}} \label{eq:conditional_beta_MNL}\\
& \text{with} \,\, \eta_{ij} = \mu_{ij} - {C}_{ij} \,\, \text{and} \,\, C_{ij} = \log \sum_{j' \neq j}^J \exp \mu_{ij'}. \notag
\end{align}
While this distribution cannot be easily sampled from, we follow the work of \cite{PolsonNicholasGandScottJamesGandWindle2013}, which has been adopted to the spatial autoregressive variant of a bivariate logit distribution \citep{Krisztin2019}. A particularly useful result in \cite{PolsonNicholasGandScottJamesGandWindle2013} is the fact that conditional on introducing a P\'{o}lya-Gamma distributed latent random variable, exponential type distributions such as the one in Eq. (\ref{eq:conditional_beta_MNL}) can be recast as Gaussian, where posterior sampling can be easily achieved. 

Particularly, when conditioning on $\omega_{ij} \sim \mathcal{PG}(1,0)$ -- where $\mathcal{PG}(1,0)$ denotes a P\'{o}lya-Gamma distribution with rate one and shape zero -- the conditional posterior of the slope parameters associated with choice $j$ can be reformulated as:
\begin{align}
 p\left(\boldsymbol{\beta}_j|\boldsymbol{\beta}_{-j}, \rho, \boldsymbol{\omega}_j\right) &\propto p\left(\boldsymbol{\beta}_j\right) \prod_{i=1}^N \exp \left( \kappa_{ij} \eta_{ij} \right) \exp\left( \frac{\eta^2_{ij} \omega_{ij}}{2} \right) \mathcal{PG}(\omega_{ij} | 1,0) \notag \\
&\propto p\left(\boldsymbol{\beta}_j\right) \exp \left\lbrace  -\frac{1}{2} \left( \left[\boldsymbol{z}_j  - \boldsymbol{c}_j\right] - \boldsymbol{A}_j^{-1} \boldsymbol{X} \right)' \boldsymbol{\Omega}_j  \left( \left[\boldsymbol{z}_j  - \boldsymbol{c}_j\right] - \boldsymbol{A}_j^{-1} \boldsymbol{X} \right) \right\rbrace, \notag
\end{align}
where $\boldsymbol{\omega}_j = [\omega_{1j},...,\omega_{Nj}]'$ and $\kappa_{ij} = y_{ij} - 1/2$. The conditional posterior has working responses $\boldsymbol{z}_j = [\kappa_{1j}/\omega_{1j},...,\kappa_{Nj}/\omega_{Nj}]'$ and $\boldsymbol{c}_j = [C_{1j},...,C_{Nj}]'$, with variance matrix $\boldsymbol{\Omega}_j = diag(\boldsymbol{\omega}_j)$. If we elicit a Gaussian prior distribution for the slope coefficients, with $p(\boldsymbol{\beta}_j) = \mathcal{N}\left( \boldsymbol{\underline{\mu}_{\beta_j}}, \boldsymbol{\underline{\Sigma}_{\beta_j}} \right)$, the conditional posteriors for the slope coefficients are also Gaussian:
\begin{align}
p\left(\boldsymbol{\beta}_j|\boldsymbol{\beta}_{-j}, \rho, \boldsymbol{\omega}_j\right) &= \mathcal{N}\left( \boldsymbol{\overline{\mu}_{\beta_j}}, \boldsymbol{\overline{\Sigma}_{\beta_j}} \right) \label{eq:beta_cond_post} \\
\boldsymbol{\overline{\mu}_{\beta_j}} &= \boldsymbol{\overline{\Sigma}_{\beta_j}} \left[ \left( \boldsymbol{A}_j^{-1}\boldsymbol{X} \right)' \left( \boldsymbol{\kappa}_j - \boldsymbol{\Omega}_j \boldsymbol{c}_j \right) +  \boldsymbol{\underline{\Sigma}_{\beta_j}}^{-1} \boldsymbol{\underline{\mu}_{\beta_j}} \right] \notag \\
\boldsymbol{\overline{\Sigma}_{\beta_j}} &= \left( \boldsymbol{A}_j^{-1}\boldsymbol{X} \right)' \boldsymbol{\Omega} \left(\boldsymbol{A}_j^{-1}\boldsymbol{X}\right) + \boldsymbol{\underline{\Sigma}_{\beta_j}}^{-1}.
\end{align}  
The Gaussian conditional posterior of the slope parameters reveals the particular appeal of using latent P\'{o}lya-Gamma distributed variables. A wide variety of Bayesian model extension, such as variable selection, or uncertainty over the $\boldsymbol{W}$ can be easily introduced in the above framework. 

Following \cite{PolsonNicholasGandScottJamesGandWindle2013}, the conditional distribution of $\boldsymbol{\omega}_j$ is also a P\'{o}lya-Gamma distribution: 
\begin{align}
p\left(\boldsymbol{\omega}_j|\boldsymbol{\beta}_1,...,\boldsymbol{\beta}_{J}, \rho_1,...,\rho_J, \boldsymbol{\omega}_{-j}\right) &= \mathcal{PG}\left(1, \boldsymbol{\eta}_j  \right), \label{eq:PG_cond_post}
\end{align}
where $\boldsymbol{\eta}_j = [\eta_1,...,\eta_N]'$. Computationally efficient algorithms for sampling from the P\'{o}lya-Gamma distribution are readily available in the R package \textbf{BayesLogit}.

The conditional posterior of $\rho$ relates directly to the multinomial logit:
\begin{align}
p\left(\rho_j| \boldsymbol{\omega}_1,...,\boldsymbol{\omega}_J,\boldsymbol{\beta}_1,...,\boldsymbol{\beta}_{J} \right) &\propto p(\rho_j) \prod_{i=1}^N \prod_{j=1}^J \frac{\left( \exp \mu_{ij} \right)^{y_{ij}}}{\sum_{j'=1}^J \exp \mu_{ij'}} \label{eq:posterior_rho} \\
\boldsymbol{\mu}_j &= \boldsymbol{A}^{-1} \boldsymbol{X} \boldsymbol{\beta}_j
\end{align}
where $p(\rho_j)$ denotes the prior distribution of $\rho_j$. The conditional posterior in Eq. (\ref{eq:posterior_rho}) is not from a well-known form and thus cannot be sampled from easily. This is usual in the spatial econometric literature, and the standard solution is to use a Metropolis-Hastings step, as in \cite{LeSage2009}.

\subsection*{Markov-chain Monte Carlo sampling procedure}

Given the conditional posterior distributions stated above, Markov-chain Monte Carlo algorithms can be employed by sequentially sampling from the conditional posteriors. We follow the usual identification assumption of the multinomial logit model in that we set $\boldsymbol{\beta}_J = 0$, $\rho_J = 0$, and $\boldsymbol{\omega}_J = 0$. With suitable starting values for $\boldsymbol{\beta}_1,...,\boldsymbol{\beta}_{J-1}$ and $\rho_1,...,\rho_{J-1}$, our sampler involves the following steps:
\begin{enumerate}[I.]
\item For $j = 1,...,J-1$, update $\boldsymbol{\omega}_j$ by drawing from $p\left(\boldsymbol{\omega}_j|\boldsymbol{\beta}_1,...,\boldsymbol{\beta}_{J}, \boldsymbol{\rho}, \boldsymbol{\omega}_{-j}\right)$ using Eq. (\ref{eq:PG_cond_post}),
\item For $j = 1,...,J-1$,update $\boldsymbol{\beta}_j$ by drawing from $p\left(\boldsymbol{\beta}_j|\boldsymbol{\beta}_{-j}, \boldsymbol{\rho}, \boldsymbol{\omega}_j\right)$ using Eq. (\ref{eq:beta_cond_post}),
\item Update $\rho_j$ using a Metropolis-Hastings step from $p\left(\rho_j| \boldsymbol{\omega}_1,...,\boldsymbol{\omega}_J,\boldsymbol{\rho}_{-j},\boldsymbol{\beta}_1,...,\boldsymbol{\beta}_{J} \right)$ based on Eq. (\ref{eq:posterior_rho}).
\end{enumerate}
The Markov-chain Monte Carlo algorithm cycles through steps I to III $B$ times by excluding the first $B_0$ draws as burn-ins. Inference on the parameters is conducted using the $B-B_0$ remaining draws.\footnote{Convergence of the MCMC algorithm was checked using the convergence diagnostics proposed by \cite{Geweke1992} and \cite{Raftery1992}. Convergence diagnostics have been calculated using the R package \textbf{coda}.}

\section{Simulation study}

In a Monte Carlo study we benchmark the SAR multinomial logit model in order to assess the predictive performance of our proposed modelling framework against two competing specifications: (i) a non-spatial version of the SAR multinomial logit, where all spatial autoregressive coefficients ${\rho}_j = 0$ for all $j$, and (ii) $J - 1$ individual SAR logit models where each logit model captures the log-odds of not choosing option $J$.\footnote{The SAR logit model is estimated using the method put forward in \cite{Krisztin2019}.} 

For the simulation study we use a SAR multinomial logit model as a benchmark data generating process, with  three choice classes ($J = 3$) and two randomly generated explanatory variables ($k=2$). The data generating process can be written as follows, where variables with a tilde denote generated quantities:
\begin{align}
\tilde{y}_{ij} &= \frac{\exp \tilde{\mu}_{ij}}{ \sum_{j'=1}^J \tilde{\mu}_{ij'}}
\end{align}
where
\begin{align}
\tilde{\boldsymbol{\mu}}_j &= \left(\boldsymbol{I} - {\rho}_j \tilde{\boldsymbol{W}} \right)^{-1} \tilde{\boldsymbol{X}} \tilde{\boldsymbol{\beta}_j} \notag \\
\tilde{\boldsymbol{\beta}}_{-J} &= \begin{pmatrix} 1 & 0.5  \\ 0.5 & 1 \end{pmatrix} + \mathcal{N}\left( \boldsymbol{0}, \begin{pmatrix} 1 & -0.25  \\ -0.25 & 1 \end{pmatrix} \right), \notag
\end{align}
$\tilde{\boldsymbol{\beta}}_{-J} = [\tilde{\boldsymbol{\beta}}_1,\tilde{\boldsymbol{\beta}}_2]$, and $\tilde{\boldsymbol{\beta}}_J = 0$. The slope coefficients and the explanatory variables are generated anew in each Monte Carlo iteration, where $\tilde{\boldsymbol{X}}$ stems from a standard normal distribution. The slope coefficients are generated as stated above. The row-stochastic spatial weight matrix $\tilde{\boldsymbol{W}}$ is based on a random spatial pattern generated from a Gaussian distribution for latitude and longitude, and constructed using seven nearest neighbours. Note, that our dependent variable $\tilde{y}_{ij}$ is a share variable, as is often used in land use share models \citep[see, e.g.][]{Chakir2009a}.

To assess the strength of the specifications along multiple scenarios, we vary the strength of spatial dependence ${{\rho}}_j \in \{0, 0.5, 0.8\}$. To evaluate the accuracy of the sampler with respect to the chosen sample size, we consider $N \in \{400, 1000\}$. Across all models, our prior set up is as follows: we use a rather uninformative Gaussian prior for $\boldsymbol{\beta}_1,..., \boldsymbol{\beta}_{J-1}$ with zero mean and variance $10^8$ and for $\rho_1,...,\rho_{J-1}$ we use a the standard beta prior specification as proposed in \cite{LeSage2009}.

\begin{threeparttable}[htbp]
  \centering
  \caption{Root mean squared error measures for the Monte Carlo runs}
  \scriptsize
\begin{tabularx}{\textwidth}{cl c*{9}{Y} G}

    \toprule
    \multirow{3}[2]{*}{$N$} & \multicolumn{1}{c}{\multirow{3}[2]{*}{Model}} & \multicolumn{9}{c}{RMSE} \\
       &    & \multicolumn{3}{c}{${{\rho}_j} = 0.0$} & \multicolumn{3}{c}{${{\rho}_j} = 0.5$} & \multicolumn{3}{c}{${{\rho}_j} = 0.8$} \\
       &    & direct & indirect & $\rho_j$ & direct & indirect & $\rho_j$ & direct & indirect & $\rho_j$ \\
    \midrule
    \multirow{3}[2]{*}{400} & SAR Multinomial logit & 0.018  & 0.024  & 0.192  & \textbf{0.018 } & \textbf{0.036 } & \textbf{0.133 } & \textbf{0.014 } & \textbf{0.067 } & \textbf{0.054 } \\
       & Multinomial logit & \textbf{0.018 } & \textbf{0.000 } & \textbf{0.000 } & 0.018  & 0.100  & 0.400  & 0.023  & 0.490  & 0.800  \\
       & Bivariate SAR logit & 0.350  & 0.158  & 0.288  & 0.348  & 0.235  & 0.150  & 0.233  & 0.559  & 0.296  \\
\midrule    
\multirow{3}[1]{*}{1,000} & SAR Multinomial logit& \textbf{0.011 } & 0.015  & 0.114  & \textbf{0.011 } & \textbf{0.021 } & \textbf{0.058 } & \textbf{0.009 } & \textbf{0.037 } & \textbf{0.015 } \\
       & Multinomial logit & 0.011  & \textbf{0.000 } & \textbf{0.000 } & 0.011  & 0.102  & 0.400  & 0.014  & 0.486  & 0.800  \\
       & Bivariate SAR Logit & 0.353 & 0.159 & 0.281 & 0.340 & 0.237 & 0.129 & 0.208 & 0.551 & 0.273 \\
\bottomrule

\end{tabularx}%
\begin{tablenotes}
\item \textbf{Notes:} Results are based on 1,000 Monte Carlo runs. For each Monte Carlo run, the corresponding sampling algorithms are run using 1,000 draws, where the initial 700 draws were discarded as burn-in. The columns \textit{direct} and \textit{indirect} correspond to summary marginal effects (for details, see the Appendix). The values given for  \textit{direct}, \textit{indirect}, and $\rho$ corresponds to the average $\text{RMSE}(\cdot)$ over all Monte Carlo iterations.  Bold values denote the lowest average $\text{RMSE}$ scores.
\end{tablenotes}
  \label{tab:MC1}%
\end{threeparttable}%
\vspace{0.5cm}

The results of the Monte Carlo study are summarized in Table \ref{tab:MC1}. Each element of the table corresponds to the average over 1,000 runs for a particular model specification and Monte Carlo scenario. The first and second columns contain information on the sample size $N$ and the model specifications. Corresponding to the choice of spatial dependence, the table reports the average root mean squared error (RMSE) point estimates for average direct and indirect impacts, as well as average estimates for $\rho_j$ for all $j$.

In the case of no spatial autocorrelation (${\rho}_j = 0$), the non-spatial multinomial logit exhibits the highest estimation accuracy for both sample sizes under scrutiny. It is worth noting that this result is hardly surprising, as in the absence of spatial autocorrelation this model resembles the true data generating process most closely.  However, that the SAR multinomial logit closely tracks the estimates of its non-spatial counterpart. In the case of $N = 1,000$, our proposed model specification even slightly outperforms all competing specifications in terms of average direct effects.

For a moderate degree of spatial autocorrelation (${\rho}_j = 0.5$), the SAR multinomial logit model outperforms all other specifications under scrutiny for both considered sample sizes. In terms of direct average impacts, the non-spatial multinomial logit model performs comparatively better. However, in the case of a smaller  sample size ($N = 400$), the bias in terms of point predictions clearly increases. Note, that the competing bivariate SAR logit specification shows considerable bias in estimating the spatial autocorrelation parameters, albeit the bias is less than that of the non-spatial multinomial logit. 

Turning attention to a high degree of spatial autocorrelation (${\rho_j} = 0.8$), we  observe that the SAR multinomial logit model significantly outperforms its alternatives. Furthermore, when high spatial autocorrelation is present, the bivariate SAR logit exhibits lower bias in terms of point prediction of average indirect effects, as the non-spatial multinomial logit model. 

Overall, we can conclude that the SAR multinomial logit model outperforms both a non-spatial multinomial logit, as well as the application of bivariate SAR logit models. This result applies both in moderate and large sample sizes. Even when no spatial autocorrelation is present, the SAR multinomial logit model produces rather promising results in terms of predictive performance, as it closely tracks the results of its non-spatial counterpart.

\section{European land use change}

Recent literature focused attention to land sealing resulting from urban sprawl, and associated spillovers to other land use classes. Results from \cite{VanVliet2019} suggest that in the last decade in Europe 8.4 Mha of land has been converted to urban, out of which 6.3 Mha was converted from cropland. However, this land sealing led to 13.1 Mha displacement of other land use classes, as cropland was expanded elsewhere, to compensate for the lack of production resources, out of which the majority (13 Mha) was expanded in other regions. These spillover effects are well documented in the literature \citep[e.g.][]{Coisnon2014,Guastella,Zoppi2014}, and serve as a motivation for an empirical application of the spatial multinomial logit model. Both global \citep{Ay2014} and local \citep{Deng2008} spillovers are considered of importance.

In the spirit of \cite{Chakir2009a}, \cite{Zoppi2014}, and \cite{Lai2017} we model the areal share of urban sprawl stemming from non-urban land in a given region within a spatial Durbin multinomial logit model, where the log odds take the following form:
\begin{align}
\boldsymbol{\mu}_j&=\rho_j\boldsymbol{W}\boldsymbol{\mu}_j+ \alpha + \boldsymbol{X}\boldsymbol{\beta}_j + \boldsymbol{W}\boldsymbol{X}\boldsymbol{\theta}_j + \boldsymbol{\varepsilon}_j. \label{eq:SDM_MNL}
\end{align}
The scalar $\alpha$ is an intercept and the term $\boldsymbol{W}\boldsymbol{X}$ is a spatial lag of the matrix of covariates with associated vector of parameters $\boldsymbol{\theta}_j$. This lag explicitly controls for the regions’ characteristics of their neighbors.

\subsection{Regions, data, and spatial weights}



Our sample covers a cross-section of 1,316 European regions across 27 countries. The regions are classified under the NUTS 2013 classification at the NUTS 3 level. They vary in size and population, however, they divide the territory of the EU for the purpose of harmonized regional statistics and analysis. Further, they are assumed to be appropriate spatial observation units for economic research and regional policy applications. The regions included in the sample are located in Austria (35 regions), Belgium (44 regions), Bulgaria (28 regions), Cyprus (one region), Czech Republic (14 regions), Denmark (eleven regions), Estonia (five regions), Finland (19 regions), France (96 regions), Germany (402 regions), Greece (52 regions), Hungary (20 regions), Italy (110 regions), Latvia (six regions), Lithuania (ten regions), Luxembourg (one region), Malta (two regions), Netherlands (40 regions), Poland (72 regions), Portugal (25 regions), Republic of Ireland (eight regions), Romania (42 regions), Slovakia (eight regions), Slovenia (twelve regions), Spain (59 regions), Sweden (21 regions), and the United Kingdom (173 regions).

The dependent variable for our analysis describes the areal share of urban sprawl emanating from any non-urban type of land within the period from 2000 to 2018. More formally, it is defined as the land area of a certain type of land use that is being transformed to urban land use between 2000 and 2018, divided by the whole area of urban expansion that took place in the respective period. As a result, we obtain a compositional data vector that -- by definition -- sums up to unity. The types of land use we consider follow the empirical literature on land use changes and urban expansion \citep{Chakir2009a,Chakir2013,Lai2016,Zoppi2014,Lai2017}. We distinguish between the five classes cropland, grassland, forest, other, and urban. At this point it is worth noting that we recognize all artificial surfaces as urban region, as we want to focus our study especially on soil sealing. The raw data stem from the CORINE Land Cover (CLC) maps provided by Copernicus Land Monitoring Service (CLMS). Their maps are based on satellite data with Minimum Mapping Units (MMU) of 25 hectares for areal phenomena and a minimum width of 100 meter for linear phenomena. The data consists of an inventory of land cover in 44 classes, which we summarize to the five classes stated above.\footnote{Table \ref{tab:clcmapping} in the appendix summarizes how each of the land cover classes was mapped.} We use CLC change-layers also provided by CLMS, designed to capture the land cover changes at a higher resolution between two neighbour surveys. Regional aggregates at the NUTS 3 level are obtained by simple summation of all changes of the corresponding raster elements. Likewise, changes for the whole investigated period are obtained by addition of the 3 sub-periods for which CLC change-layers are provided. Further data sources are i) the Urban Data Platform Plus provided as a joint initiative of the Joint Research Centre (JRC) and the Directorate General for Regional and Urban Policy (DG REGIO) of the European Commission, ii) Eurostat (the statistical office of the European Union) and iii) the European Observation Network for Territoral Development and Cohesion (ESPON).

\begin{table}[h!]
\setlength{\extrarowheight}{2.5pt}
  \centering
  \caption{Variables used in the empirical analysis}
    \begin{adjustbox}{width=\textwidth}
    \begin{tabular}{lp{26.5em}l}
    \toprule
    \toprule
    \textbf{Variable } & \multicolumn{1}{l}{\textbf{Description}} & \textbf{Source} \\
    \midrule
    \multicolumn{1}{p{8.72em}}{Cropland \newline{}to artificial} & Sum of 2000-2006, 2006-2012, and 2012-2018 CLC land-cover \newline{}changes from cropland to artificial land, divided by \newline{}the total change of artificial area in the same period. & CLC \\
    \multicolumn{1}{p{8.72em}}{Forest\newline{}to artificial} & Sum of 2000-2006, 2006-2012, and 2012-2018 CLC land-cover \newline{}changes from forests to artificial land, divided by \newline{}the total change of artificial area in the same period. & CLC \\
    \multicolumn{1}{p{8.72em}}{Grassland \newline{}to artificial} & Sum of 2000-2006, 2006-2012, and 2012-2018 CLC land-cover \newline{}changes from pastures and grassland to artificial land, divided \newline{}by the total change of artificial area in the same period. & CLC \\
    \multicolumn{1}{p{8.72em}}{Other\newline{}to artificial} & Sum of 2000-2006, 2006-2012, and 2012-2018 CLC land-cover \newline{}changes from area of other use to artificial land, divided by \newline{}the total change of artificial area in the same period. & CLC \\
    \midrule
    Crop rent & Share of agricultural gross value added, divided by square km \newline{}of area used to grow crops, 2000. & JRC, CLC \\
    Forest rent & Share of agricultural gross value added, divided by square km \newline{}of forest-area, 2000. & JRC, CLC \\
    Grass rent & Share of agricultural gross value added, divided by square km \newline{}of pasture and grassland, 2000. & JRC, CLC \\
    Initial artificial area & \multicolumn{1}{l}{Area of artificial land cover, 2000.} & CLC \\
    Artificial growth & Growth of artificial areas between 2000 and 2018 \newline{}measured in percent. & CLC \\
    Employment primary & Share of employment in the primary sector (NACE A), in total \newline{}employment, 2000. & JRC \\
    Employment tertiary  & Share of employment in the tertiary sector (NACE F to Q) in \newline{}total employment, 2000. & JRC \\
    Gdp per capita & \multicolumn{1}{l}{Gross domestic product divided by population,  2000.} & JRC \\
    Population density & \multicolumn{1}{l}{Population per square km, 2000.} & JRC \\
    Elevation & \multicolumn{1}{l}{Average elevation in meters.} & Copernicus \\
    Slope & \multicolumn{1}{l}{Average slope in degree.} & Copernicus \\
    Soil moisture & Content of liquid water in a surface soil layer of 2 to 5 cm depth \newline{}expressed as qubic m water per qubic m of soil, 2000. & Copernicus \\
    N2000 cropland & Share of protected area used to grow crops over total area used \newline{}to grow crops, 2000. & Natura 2000 \\
    N2000 forest & \multicolumn{1}{l}{Share of protected area of forests over total area of forests, 2000.} & Natura 2000 \\
    N2000 grassland  & Share of protected area of pastures and grassland over total area \newline{}of pastures and grassland, 2000. & Natura 2000 \\
    N2000 other & Share of protected area of other use over total area \newline{}of other use, 2000. & Natura 2000\\
    Objective 2 region & Dummy varible, 1 denotes region eligible under objective 2 \newline{}2000–2006, 0 otherwise. & ESPON \\
    Farm density & Number of farms divided square km, measured \newline{}on NUTS2 level, 2000. & Eurostat \\
    Farm size & Total farm area devided by number of farms, measured \newline{}on NUTS2 level, 2000. & Eurostat \\
    \bottomrule
    \bottomrule
    \end{tabular}%
      \end{adjustbox}
  \label{tab:data}%
\end{table}%

Our set of covariates consists of $K^\prime=19$ candidate variables that are commonly employed in the literature on land use changes \citep[for an overview, see][]{Shaw2020}. Further, to capture the complex spatial structure we include not only the spatially lagged dependent vector, but also the spatially lagged forms of the explanatory variables (except for the dummy variables). We also include a vector of ones as intercept. Therefore, the resulting design matrix is of column-dimension $K=K^\prime+18+1=38$ where 18 are the spatially lagged covariates and 1 is the intercept. Table \ref{tab:data} provides a short technical description for the variables included in our estimation.

Since the rent of a certain land use class is assumed to affect the decision of land-owners -- yet it is usually not observed -- many recent studies consider various proxies to control for the variation in returns from different land uses \citep[see, e.g.][]{Fan2005,Lubowski2008}. \cite{Chakir2009a} conclude that agricultural gross value added divided by the respective land use area serves as a reasonably good proxy. Higher rents are therefore assumed to reduce the amount of land that is converted to artificial area.

The initial level of artificial areas and -- especially -- urban expansion rates are discussed in the literature in the context of the level of available agricultural amenities \citep{Wu2006,Wu2003,Coisnon2014}. Based on this strain of literature lower initial urban expansion would lead to higher urbanization rates, as regions surrounding population centres with low urban share are in higher demand.

On the other hand, quantities on employment, population and income are typical variables to represent the degree of economic development. Employment enters the model in the form of sectoral shares, with manufacturing (secondary) as baseline. Region specific population, a particularly important driver of land take \citep[see e.g.][]{Guastella,Terama2019,Paulsen2012}, is divided by the respective area and therefore captured as density. Income is measured as gross domestic product per inhabitant. High shares of tertiary employment, paired with high income and population density is usually observed around the city centres and, therefore, associated with expansion of housing supply which again should translate into urban expansion.

Quantities usually associated with the quality of soil include measures of slope, elevation and moisture (usually in form of precipitation or humidity). Following \cite{Chang-Martinez2015} we include these physical drivers of land use conversion, as they implicitly influence the cost of land conversion. We consider slope and elevation in average meters and degrees respectively. Soil moisture is captured as volumetric measure of liquid water in a surface soil layer of 2 to 5 cm depth. Variables capturing the quality of the land are assumed to have a negative impact on conversion of productive land, as they are to be interpreted as costs of conversion \citep{Shaw2020,Gruenberg2006}. 

Additionally, national regulations, as the amount of nature conservation areas, restrict the potential conversion. We include the share of area being protected under the Natura 2000 network of nature protection. The Natura 2000 network's main objective is to preserve natural habitats and secure biodiversity in the European Union, hence, forest and grassland areas are of main concern \citep{Lai2017}. 

In the discussion of steering soil sealing, subsidies and taxes play a key role \citep{Artmann2014a,Shaw2020}. As a proxy for European level subsidies we utilize observation on whether a region received Objective 2 level regional funding within the period, as this type of funding is also used to enhance infrastructure in the region. An additional major source of subsidy for land use management are agricultural subsidies of countries, as well as the European Union. These are not divided on the regional level, but by farm size and productivity. Therefore, to control for the heterogeneous structures of agricultural actors across Europe, variables that account for farm specific characteristics are incorporated \citep[for a discussion see][]{Delbecq2014}. 

For the spatial weights matrix $\boldsymbol{W}$ we suppose a neighbourhood structure known as seven nearest neighbour specification, where every region is constrained to be a neighbour of its seven closest regions. Our results, however, prove robust to variations in the assumed spatial dependence structure.

\subsection{Empirical results}
This subsection presents the Markov Chain Monte Carlo (MCMC) results obtained from 10,000 posterior draws for our spatial multinomial logit specification, where the first 5,000 were discarded as burn-in.\footnote{Convergence of the sampler was checked using the diagnostics by \citep{Geweke1992}} Straightforward interpretation of coefficient estimates in spatial models could lead to deceptive or misleading conclusions \citep[see, e.g. ][]{Anselin1988,LesageFischer2009}. One possibility is to provide summary metrics in form of direct, indirect (spillover) and total effects. Following \citep{LeSage2009} we present marginal effects in Table \ref{tab:res}. Direct effects are then to be interpreted similar to regular slope coefficients. In turn, indirect effects account for the impacts due to changes in other regions and are therefore to be interpreted as spillover effects. We find a significant class-specific spatial parameter $\rho_j$ for cropland as well as grassland, highlighting the necessity of incorporating the spatial dependence structure in the model. This result confirms the findings of \cite{Guastella} and especially of \cite{VanVliet2019}, in that the expansion of artificial surfaces on productive land leads to further spillover land conversions in surrounding regions.

In addition, the table reports the McFadden pseudo $R\textsuperscript{2}$, which serves as a measure of the goodness of fit in limited dependent variable models.  \cite{McFadden1974} highlights that values between 0.2 and 0.4 already indicate a rather good fit. The rest of the reported results is to be interpreted as follows: For a certain explanatory variable the four values reported in the direct effect column represent the class specific (cropland, forest, grassland, and other) responses to variation in the respective explanatory. These responses are the changes of the probabilities to convert the respective class in that region to artificial area. Similar, the four class specific columns for the indirect effect are the responses to changes in the explanatory variable in all other regions. 


\begin{table}[h!]
\setlength{\extrarowheight}{3pt}
  \centering
  \caption{Summary impact measures for artificial area expansion from each land use class}
      \begin{adjustbox}{width=\textwidth}
    \begin{tabular}{lrrrrrrrr}
    \toprule
    \toprule
          & \multicolumn{4}{c}{Direct}   & \multicolumn{4}{c}{Indirect} \\
        \cmidrule(lr){2-5}\cmidrule(lr){6-9}
        & \multicolumn{1}{c}{Cropland} & \multicolumn{1}{c}{Forest} & \multicolumn{1}{c}{Grass} & \multicolumn{1}{c}{Other} & \multicolumn{1}{c}{Cropland} & \multicolumn{1}{c}{Forest} & \multicolumn{1}{c}{Grass} & \multicolumn{1}{c}{Other} \\
    \midrule
    Crop rent & \textbf{-0.046} & 0.017 & \textbf{0.031} & \multicolumn{1}{r}{-0.003} & -0.055 & 0.042 & -0.002 & 0.012 \\
    Forest rent & \textbf{0.084} & \textbf{-0.061} & -0.008 & \multicolumn{1}{r}{-0.014} & 0.021 & -0.038 & 0.008 & 0.010 \\
    Grass rent & \textbf{0.048} & -0.011 & \textbf{-0.031} & \multicolumn{1}{r}{-0.006} & -0.023 & 0.004 & 0.008 & 0.012 \\
    Initial artificial area & -0.041 & -0.003 & 0.008 & \multicolumn{1}{r}{\textbf{0.035}} & 0.009 & 0.010 & -0.033 & 0.015 \\
    Artificial growth & -0.008 & -0.007 & 0.007 & \multicolumn{1}{r}{0.007} & 0.022 & -0.012 & 0.022 & -0.033 \\
    Employment primary & -0.015 & -0.012 & 0.018 & \multicolumn{1}{r}{0.009} & -0.029 & -0.011 & \textbf{0.067} & -0.019 \\
    Employment tertiary & -0.048 & 0.002 & 0.013 & \multicolumn{1}{r}{\textbf{0.035}} & -0.056 & -0.019 & \textbf{0.084} & -0.005 \\
    Gdp per capita & 0.022 & 0.024 & \textbf{-0.065} & \multicolumn{1}{r}{0.018} & 0.052 & -0.016 & -0.020 & -0.021 \\
    Population density & 0.037 & \textbf{-0.034} & \textbf{0.046} & \multicolumn{1}{r}{\textbf{-0.050}} & -0.039 & -0.015 & 0.037 & 0.021 \\
    Elevation & 0.006 & -0.017 & 0.020 & \multicolumn{1}{r}{-0.005} & 0.014 & -0.006 & 0.006 & -0.017 \\
    Slope & -0.042 & 0.021 & -0.003 & \multicolumn{1}{r}{\textbf{0.023}} & -0.002 & 0.012 & -0.051 & 0.034 \\
    Soil moisture & -0.012 & 0.009 & 0.010 & \multicolumn{1}{r}{-0.005} & -0.043 & -0.004 & \textbf{0.077} & \textbf{-0.027} \\
    N2000 cropland & -0.015 & 0.009 & 0.007 & \multicolumn{1}{r}{-0.001} & -0.062 & 0.007 & 0.021 & 0.036 \\
    N2000 forest & \textbf{0.052} & \textbf{-0.027} & -0.019 & \multicolumn{1}{r}{-0.007} & \textbf{0.060} & -0.011 & \textbf{-0.059} & 0.009 \\
    N2000 grassland & 0.012 & 0.017 & \textbf{-0.029} & \multicolumn{1}{r}{0.004} & 0.043 & -0.009 & -0.006 & -0.028 \\
    N2000 other & -0.009 & 0.003 & 0.017 & \multicolumn{1}{r}{-0.011} & 0.051 & -0.014 & 0.017 & \textbf{-0.052} \\
    Objective 2 region & \textbf{-0.108} & 0.041 & -0.016 & \multicolumn{1}{r}{\textbf{0.070}} & \textbf{-0.006} & 0.000 & -0.001 & 0.000 \\
    Farm density & 0.020 & 0.020 & \textbf{-0.040} & \multicolumn{1}{r}{0.003} & -0.018 & -0.008 & 0.039 & -0.006 \\
    Farm size & -0.010 & -0.005 & 0.013 & \multicolumn{1}{r}{0.003} & -0.025 & 0.008 & 0.004 & 0.011 \\
    \midrule
    $\rho_j$       & \textbf{0.067} & 0.017 & \textbf{0.082} & 0.000 &      &       &       &    \\
    McFadden $R^2$ & \textbf{0.119} &       &       &       &       &       &       &  \\
    \bottomrule
    \bottomrule
    \end{tabular}%
          \end{adjustbox}
          \vspace*{-0.4cm}
          \begin{flushleft}
            \textbf{Notes}: Summary metrics are based on 10,000 Markov Chain Monte Carlo iterations, where the first 5,000 were discarded as burn-in. Bold written estimates indicate statistical significance under a 90\% credible interval.
    
          \end{flushleft}

  \label{tab:res}%
\end{table}%

The direct effects of the three types of land rent proxies (crop, forest and grass) confirm results from \cite{Chakir2017} and \cite{Chakir2009a}, in that for each land use class higher rents imply a significantly lower chance of conversion. Additionally, the joint modelling in a multinomial model indicates that significant spillover effects to other classes are present. Most notably, an increase in cropland rents in a region, would also increase the conversion of grassland to artificial areas. We find analogous relationships for forest rent and cropland, as well as grass rent and cropland.

A higher initial level of artificial areas indicates that land sealing of other natural vegetation in the own region has a significantly higher probability as compared to land sealing of the other land covers under scrutiny. \cite{Burnett2012} have similar findings, where urbanization is a process which enforces itself. Moreover, as the crop, grass, and forest land surrounding cities is frequently the most productive \citep{Shaw2020}, it seems intuitive that urban expansion would take from the comparatively less productive other natural vegetation.The change of artificial area appears to have no direct or indirect impacts on the allocation of its origin.

Regarding the sectoral mix of employment, our results indicate that a higher share of tertiary employment in the own region implies a significantly higher probability of other natural vegetation being converted to artificial land. This reflects the findings of \cite{Salvati2016a} and \cite{Salvati2016}, where higher tertiary employment is found to mainly reflect the presence of urban fabric. In this context, the positive spillover effects of primary and tertiary employment to neighbouring regions' grassland can be contextualized as the effect of industrial belts on  pastures. This reflects the findings of the theoretical model of \cite{Turner2005}, where chiefly industry clusters agglomerate with large-scale livestock farms.

Our results with regards to gross domestic product per capita suggest that it is not a significant driver of land sealing in a European context. This is as opposed to findings of e.g. \cite{Deng2008} in developing countries, where gdp per capita is found to be one of the main drivers of urbanization. Moreover, we find that a higher gdp per capita in fact significantly lowers the probability of sealing grass land in the own region. When observed jointly with the direct effects of population density, this supports findings by \cite{McGrath2005} and \cite{Guiling2009}, who find that population is a more significant driver of urbanization, as opposed to personal income. Note, however that only for the land take from grass land is the effect positive and significant. For land takes from forest and other natural land, a higher population density in fact results in a lower chance of land conversion. This result can be interpreted on the one hand with the fact that regions with a higher endowment of population density are more urban in nature and contain a much lower percentage of cropland or other natural vegetation. On the other hand, specific literature, such as \cite{Delbecq2014} and \cite{Wu2006}, provide evidence that private home-owners exhibit strong preferences for surrounding grassland amenities.

Turning our attention to the estimated impacts of the biophysical drivers elevation, slope, and soil moisture, we can largely confirm the overall conclusions of \cite{Shaw2020} and \cite{Chang-Martinez2015} in that the biophysical processes play a secondary role to socio-economic ones in explaining land sealing processes. For the own-region, only slope plays has a small, albeit significant impact on the probability of sealing other natural vegetation. Additionally, a higher percentage of soil moisture indicates a significantly higher chance of converting grassland to urban land in neighbouring regions.

Our results seem to indicate that the \textit{Natura 2000} protection program has intended effects, as higher shares of protected forest and grassland would -- according to our results -- lead to significantly lower chance of the respective land cover being converted into urban. Note, that if a region has a higher share of other natural vegetation under Natura 2000 protection, this would lower the chances of neighbouring regions converting this land cover to urban. This largely confirms the findings of \cite{Lai2016}, \cite{Zoppi2014}, and \cite{Lai2017}. Our joint multinomial logit framework, however, allows us to uncover additional interdependencies among the natural protection of land covers. Our estimated results suggest that a higher share of protected forest in a NUTS3 level region would results in a significantly higher probability of converting cropland to urban, not only in the own region but also amongst neighbours. Additionally, this would also lower the odds sealing grassland under artificial surfaces.

The estimated results with regard to our subsidy proxies seem to show that regional funding plays a comparatively larger role as farm-specific subsidies. The own-regional effect of regional level Objective 2 subsidies is significant and negative for cropland, and positive for other natural vegetation. This finding supports the hypotheses that subsidies increase land conversion \citep{Shaw2020}. Particularly noteworthy is the result that the land take comes more significantly from natural vegetation (which is highest in biodiversity) as opposed to more productive cropland. Additionally, neighbours of regions under Objective 2 funding also have a significantly decreased chance of converting cropland to artificial areas. This might suggest an increase in cropland productivity, through better infrastructure. With regard to our farm structure variables -- proxying the role of the Common Agricultural Policy -- our results indicate a significant effect only with regard to farm density: a higher density of farms would lower the chance of converting land to grassland in the own region.

\section{Concluding remarks}

In this paper we put forth a Bayesian estimation approach for a multinomial logit specification for the modelling of land use conversion, which has a spatial autoregressive structure in the log odds, with differing strength of spatial autocorrelation for each choice alternative. The virtue of our specification is that it combines a spatial autoregressive framework (allowing for cross regional spillovers), and a joint multinomial framework (allowing for cross land use class dependencies). The proposed approach is based on recent spatial econometric advances dealing with Bayesian estimation of the logit model \cite{Krisztin2019}. The core step of the estimation procedure relies on introducing  a latent P\'{o}lya-Gamma variable \citep[see][]{PolsonNicholasGandScottJamesGandWindle2013}. Through the latent variable, the conditional posterior distribution of the slope parameters in the spatial autoregressive logit specification is rendered in  a Gaussian form, which allows us to tackle the  MCMC estimation in a particularly efficient way. We demonstrate in a simulation study the advantages and behaviour of our proposed model specification, benchmarking it against simpler alternatives.

The virtues of the spatial multinomial logit model are illustrated using an empirical specification. Specifically, we examine the land sealing activities in European NUTS-3 level regions. We consider the areal share of urban sprawl emanating from cropland, grassland, forest, and other natural vegetation from 2000 to 2018. The observation on land use data stem from the CORINE Land Cover (CLC) maps. Our results suggest, that spatial dependence indeed play a small, but significant role, particularly for the land use classes cropland and grassland. For all land covers proxied land rents are of central importance. Additionally, our findings corroborate evidence from recent literature, that socio-economic drivers play a much more central role, as opposed to biophysical ones \citep[for an overview, see][]{Shaw2020}. The key role of population density \cite{Guastella,Deng2008,Lai2016} in urban land take is confirmed by our results.  Moreover, we confirm on a larger level that environmental protection not only has effects in the own- but also in neighbouring regions \citep{Lai2016,Zoppi2014,Lai2017}. Through the virtue of our multinomial analysis we also find evidence for forest protection having spillover effects to neighbouring regions and other land covers.

\bibliographystyle{fischer}
\bibliography{library,MiscBib}

\newpage

\appendix

\renewcommand\thesection{\Alph{section}}
\setcounter{section}{1}

\renewcommand{\thetable}{A\arabic{table}}
\renewcommand{\thefigure}{A\arabic{figure}}
\setcounter{figure}{0}
\setcounter{table}{0}

\section*{Appendix}

\begin{table}[htb]
\setlength{\extrarowheight}{-0.5pt}
  \centering
  \caption{Mapping of detailed CLC classes to land use aggregates}
    \begin{adjustbox}{width=\textwidth}
    \begin{tabular}{cp{26.5em}l}
    \toprule
    \toprule
    \textbf{CLC code} & \multicolumn{1}{l}{\textbf{Description}} & \textbf{Aggregate} \\
    \midrule
    111 & Continuous urban fabric & Artificial\\
112 & Discontinuous urban fabric & Artificial\\
121 & Industrial or commercial units & Artificial\\
122 & Road and rail networks and associated land & Artificial\\
123 & Port areas & Artificial\\
124 & Airports & Artificial\\
131 & Mineral extraction sites & Artificial\\
132 & Dump sites&  Artificial\\
133 & Construction sites&  Artificial\\
141 & Green urban areas & Artificial\\
142 & Sport and leisure facilities & Artificial\\
211 & Non-irrigated arable land & Cropland\\
212 & Permanently irrigated land & Cropland\\
213 & Rice fields & Cropland\\
221 & Vineyards & Cropland\\
222 & Fruit trees and berry plantations & Cropland\\
223 & Olive groves & Cropland\\
231 & Pastures & Grassland\\
241 & Annual crops associated with permanent crops & Cropland\\
242 & Complex cultivation patterns & Cropland\\
243 & Land principally occupied by agriculture \& natural vegetation & Forest\\
244 & Agro-forestry areas & Forest\\
311 & Broad-leaved forest & Forest \\
312 & Coniferous forest & Forest\\
313 & Mixed forest & Forest\\
321 & Natural grasslands & Other\\
322 & Moors and heathland & Other\\
323 & Sclerophyllous vegetation & Other\\
324 & Transitional woodland-shrub & Other\\
331 & Beaches dunes sands & Other\\
332 & Bare rocks & Other\\
333 & Sparsely vegetated areas & Other \\
334 & Burnt areas & Other\\
335 & Glaciers and perpetual snow & excluded\\
415 & Inland marshes & excluded\\
412 & Peat bogs & excluded\\
421 & Salt marshes & excluded\\
422 & Salines & excluded\\
423 & Intertidal flats  & excluded\\
511 & Water courses & excluded\\
512 & Water bodies & excluded\\
521 & Coastal lagoons & excluded\\
522 & Estuaries & excluded\\
523 & Sea and ocean & excluded\\
999 & NODATA & excluded\\
    \bottomrule
    \bottomrule
    \end{tabular}%
      \end{adjustbox}
  \label{tab:clcmapping}%
\end{table}%

\subsection*{Marginal effects}

Similar to the marginal effects of the spatial Durbin logit model \citep[see][]{Krisztin2019}, in the presented multinomial logit model in Eq. (\ref{eq:SDM_MNL}) the interpretation of  marginal effects of the $k$-th explanatory variable (with $k = 1,...,K$) differs from those in linear models. This is due to the fact that the multinomial logit model is non-linear in nature, but also the the presence of spatial autocorrelation gives rise to an $N \times N$ matrix of partial derivatives, which makes interpretation of marginal effects richer, but also more complicated (see also \citealt{LeSage2009}).

As is standard in the logit literature, and analogous with the proposed marginal effects of the spatial logit model \citep{Krisztin2019}, we provide marginal effects relative to the mean of the $k$-th explanatory variable, which we denote as $\overline{x_k} = \sum_{i=1}^N x_{ik} /N$. Thus the interpretation of the marginal effects is the change in probability of observing $y=j$ associated with a change in the average sample observation of the $k$-th explanatory variable. To write the partial derivatives of the model in Eq. (\ref{eq:SDM_MNL}), with respect to the $k$-th coefficient let us define:
\begin{align*}
\boldsymbol{\mu}_{kj} &= \boldsymbol{A}_j^{-1} \boldsymbol{I}_N \overline{x_k} \beta_{kj} + \boldsymbol{A}_j^{-1} \boldsymbol{W} \overline{x_{Wk}} \theta_{kj}, \notag \\
\boldsymbol{\zeta}_{kj} &= \boldsymbol{A}_j{-1} \boldsymbol{I}_N \beta_{kj} + \boldsymbol{A}_j^{-1} \boldsymbol{W} \theta_{kj}, \,\, \text{and} \notag \\
\boldsymbol{p}_{kj} &= \frac{\exp\boldsymbol{\mu}_{kj}}{\sum^J_{j'} \exp\left(\boldsymbol{\mu}_{kj'} \right)}. 
\end{align*}
$\beta_{kj}$ and $\theta_{kj}$ denote the $k$-th element of $\boldsymbol{\beta}_j$ and $\boldsymbol{\theta}_j$, respectively. $\overline{x_{Wk}}$ denotes the average value of the $k$-th spatially lagged explanatory variable. The partial derivatives can then be expressed as:
\begin{align}
\frac{\partial p({y}=j|\overline{x}_k )}{\partial \overline{x}_k'} &= \boldsymbol{p}_{kj}  \odot \left[ \boldsymbol{\zeta}_{kj} - \sum^J_{j'} \boldsymbol{p}_{kj'} \odot \boldsymbol{\zeta}_{kj'} \right]  \label{eq:marg_fx_logit}, \\
&= \boldsymbol{\Lambda}_{kj}, \notag
\end{align}
where $\odot$ is the Hadamard product. Note that marginal effects of the $k$-th coefficient on class $j$, denoted as $\boldsymbol{\Lambda}_{kj}$, are an $N\times N$ matrix due to the presence of the $N \times N$ spatial multiplier $\boldsymbol{A}_j^{-1}$.

Interpreting  $N \times N$ marginal effects proves cumbersome, therefore we define summary impact effects \citep{LeSage2009}. These can be readily calculated from $\boldsymbol{\Lambda}_{kj}$:
\begin{align}
direct_{kj} &= \frac{1}{N}\boldsymbol{\iota}_N' diag(\boldsymbol{\Lambda}_{kj}) \\
total_{kj} &= \frac{1}{N}\boldsymbol{\iota}_N' \boldsymbol{\Lambda}_{kj} \boldsymbol{\iota}_N  \\
indirect_{kj} &= total_{kj} - direct_{kj}, 
\end{align}
where $\boldsymbol{\iota}_N$ denotes an $N \times 1$ vector of ones. 


\end{document}